# Amendment to "Performance Analysis of the V-BLAST Algorithm: An Analytical Approach." [1]

Sergey Loyka
School of Information Technology and
Engineering (SITE), University of Ottawa,
161 Louis Pasteur, Ottawa, Ontario,
Canada, K1N 6N5
E-mail: sergey.loyka@ieee.org

Francois Gagnon
Department of Electrical Engineering
Ecole de Technologie Superieure
1100, Notre-Dame St. West, Montreal
Quebec, H3C 1K3, Canada
E-mail: francois.gagnon@etsmtl.ca

An analytical technique for the outage and BER analysis of the nx2 V-BLAST algorithm with the optimal ordering has been presented in [1], including closed-form exact expressions for average BER and outage probabilities, and simple high-SNR approximations. The analysis in [1] is based on the following essential approximations:

1. The SNR was defined in terms of total after-projection signal and noise powers, and the BER was analyzed based on their ratio. This corresponds to a non-coherent (power-wise) equal-gain combining of both the signal and the noise, and it is not optimum since it does not provide the maximum output SNR.

2. The definition of the total after-projection noise power at each step ignored the fact that the after-projection noise vector had correlated components.

3. The after-combining noises at different steps (and hence the errors) were implicitly assumed to be independent of each other. Under non-coherent equal-gain combining, that is not the case.

It turns out that the results in [1] hold also true without these approximations, subject to minor modifications only. The purpose of this note is to show this and also to extend the average BER results in [1] to the case of BPSK-modulated V-BLAST with more than two Rx antennas (eq. 18-20). Additionally, we emphasize that the block error rate is dominated by the first step BER at the high-SNR mode (eq. 14 and 21).

To accomplish the task, we employ the maximum ratio combining (MRC) weights, which also incorporate the orthogonal projection to eliminate interference from yet-to-be-detected symbols (i.e. zero-forcing (ZF) MRC weights, which maximize the SNR under the ZF constraint), and the expressions for the corresponding after-processing SNR. For such weights, the after-combining noises at different steps (and hence errors) are indeed independent of each other. Based on this, we demonstrate that the results in [1] apply to the case of the ZF-MRC processing with minor modifications only: the exact closed-form expressions for the average BER hold true up to a constant factor in terms of the average SNR.

The standard baseband system model is given by [1, eq. 1 and 2]. The received signal after the interference cancellation at the i-th step $\mathbf{r}_i'$ is given by [1, eq. 3]. The inter-stream interference nulling (from yet-to-be-detected symbols) can be expressed as

$$\mathbf{r}_i'' = \mathbf{P}_i \mathbf{r}_i' = \mathbf{P}_i \mathbf{h}_i q_i + \mathbf{P}_i \mathbf{v} \qquad (1)$$

where $\mathbf{P}_i$ is the projection matrix onto the sub-space orthogonal to $span\{\mathbf{h}_{i+1}, \mathbf{h}_{i+2}, ... \mathbf{h}_m\}$: $\mathbf{P}_i = \mathbf{I} - \mathbf{H}_i (\mathbf{H}_i^+ \mathbf{H}_i)^{-1} \mathbf{H}_i^+$, where $\mathbf{H}_i = [\mathbf{h}_{i+1} \mathbf{h}_{i+2} ... \mathbf{h}_m]$ and "+" denotes conjugate transpose. The first term in (1) represents the signal and the second one is the noise contribution. The analysis in [1] was based on the after-processing signal and noise powers defined as $P_{si} = |\mathbf{h}_{i\perp}|^2$ (assuming that $|q_i|^2 = 1$), where $\mathbf{h}_{i\perp} = \mathbf{P}_i \mathbf{h}_i$, and $P_{vi} = \langle |\mathbf{P}_i \mathbf{v}|^2 \rangle_\mathbf{v} = (n-m+i)\sigma_0^2$, where $\langle \ \rangle_\mathbf{v}$ is the expectation over the noise. Consequently, the output SNR was obtained as

$$\gamma_i' = \frac{P_{si}}{P_{vi}} = \frac{|\mathbf{h}_{i\perp}|^2}{(n-m+i)\sigma_0^2} \qquad (2)$$

and the outage and BER analysis was carried out in terms of $\gamma_i'$ [1]. This definition of the output SNR corresponds to power-wise (non-coherent) combining with unit weights (i.e. equal gain) in terms of $\mathbf{r}_i''$, which is not optimum. Additionally, the definition above ignores the fact that the after-projection noise in (1), $\mathbf{v}_i' = \mathbf{P}_i \mathbf{v}$, has correlated components [2]: $\mathbf{R}_{\mathbf{v}i} = \langle \mathbf{v}_i'(\mathbf{v}_i')^+ \rangle = \sigma_0^2 \mathbf{P}_i \neq \sigma_0^2 \mathbf{I}$.

*Optimum Weights*: To remove these approximations, we employ the optimum combining weights that maximize the output after-projection SNR in (1) and maintain orthogonality to $\mathbf{H}_i = [\mathbf{h}_{i+1} \mathbf{h}_{i+2} ... \mathbf{h}_m]$ (ZF-MRC weights) [2,3],

$$\mathbf{w}_i = \frac{\mathbf{h}_{i\perp}}{|\mathbf{h}_{i\perp}|} \tag{3}$$

Similar approach has been used before in the context of multiuser CDMA receivers [4,5]. Using the ZF-MRC weights in (4), the output SNR becomes [2,3],

$$\gamma_i = \frac{|\mathbf{h}_{i\perp}|^2}{\sigma_0^2} \tag{4}$$

Comparing (4) to (2), one concludes that the optimum SNR has the same distribution (up to a constant factor) as the "power-wise" one, $\gamma_i = (n - m + i)\gamma_i'$. Hence, all the results in [1] in terms of the SNR distribution can be applied to the ZF-MRC processing with a minor modification, i.e. the substitution $\gamma_i' \to \gamma_i$.

*Independence of the After-Combining Noises*: Using the following property of the projection matrices, $\mathbf{P}_i \mathbf{P}_j = \mathbf{P}_i, \forall j > i$, it follows that the weights are orthogonal to each other: $\mathbf{w}_i^+ \mathbf{w}_j = \mathbf{h}_i^+ \mathbf{P}_i \mathbf{P}_j \mathbf{h}_j = 0 \ \forall i \neq j$. The after-combining noise correlation at step $i$ and $j$ is $R_{ij} = \langle \xi_i^* \xi_j \rangle = \sigma_0^2 \mathbf{w}_i^+ \mathbf{w}_j = \sigma_0^2 \delta_{ij}$, where $\xi_i = \mathbf{w}_i^+ \mathbf{v}$ is the after-combining noise, the expectation is over the noise, and $\delta_{ij} = 1$ if $i = j$ and 0 otherwise. Since $\xi_i$ are complex Gaussian (for a given channel realization), independence follows from zero correlation. This result parallels one for multi-user receivers [4,5]. Since the after-combining noises are i.i.d. and also independent of the channel, the decision statistics and errors are independent at each step[1], which simplifies the analysis significantly. This was implicitly assumed in [1][2].

*Outage probability and average BER in a Rayleigh fading channel*: Based on the discussion above, we conclude that all the results in [1] on the outage probability, which were originally formulated in terms of the signal power, also hold in terms of the optimum SNR, as (4) demonstrates. Using [1, eq.25 and 29] and after some manipulations, the 1st and 2nd step outage probabilities (or the SNR CDF) can be expressed in the following form, which emphasizes the link with the MRC outage probability and also facilitates the average BER analysis,

$$F_1(x) = 2F_{MRC}^{(n-1)}(x) - F_{MRC}^{(n-1)}(2x) + \Delta F_1(x) \qquad (5)$$

$$F_2(x) = F_{MRC}^{(n)}(x)\left[2 - F_{MRC}^{(n)}(x)\right] = F_{MRC}^{(n)}(2x) - e^{-2x}\sum_{k=n}^{2n-2} b_k x^k \qquad (6)$$

where $F_i(x) = \Pr\{\gamma_i / \gamma_0 < x\}$, $\gamma_0$ is the average SNR, $F_{MRC}^{(n)}(x)$ is the n-th order MRC outage probability,

$$F_{MRC}^{(n)}(x) = 1 - e^{-x}\sum_{k=0}^{n-1}\frac{x^k}{k!} \qquad (7)$$

and

$$\Delta F_1(x) = e^{-2x}\sum_{i=n-1}^{2n-3} a_i (2x)^i,$$

$$a_i = \frac{n-1}{(i-n+1)!}\sum_{j=i+1}^{2n-2}\frac{(j-n)!}{2^j}\sum_{k=j-n+1}^{n-1}\frac{1}{k!(j-k)!} \qquad (8)$$

$$b_k = \sum_{i=k-n+1}^{n-1}\frac{1}{i!(k-i)!}$$

They asymptotic behavior (for large average SNR) is given by [1, eq. (26) and (33)]. Corresponding PDFs can be obtained by differentiation, $\rho_i(\gamma) = dF_i(\gamma/\gamma_0)/d\gamma$. Eq. (36) in [1] gives the instantaneous block error rate (BLER), i.e. the probability that there is at least one error in the decoded Tx vector symbol (which was termed "total instantaneous probability of error" in [1]), and it already takes into account the error independence at each step, which was assumed without proof in [1] and which was rigorously demonstrated above. The average BLER, which was termed in [1] "total average BER", is given by in [1, eq. 37]. Using the optimum SNR result in (4), the average BER expressions in [1] have to be modified as follows. For non-coherent binary orthogonal FSK, $P_e(\gamma) = e^{-\gamma/2}/2$ and eq. 39 in [1] holds true; [1, eq. 40] should be modified to,

$$\overline{P_{e,2}} = \frac{4}{(4+\gamma_0)^2} + \frac{16}{(4+\gamma_0)^3} \tag{9}$$

---

[1] assuming no error propagation, which is sufficient for block error rate analysis
[2] it should be pointed out that this independence holds true under the ZF MRC weights, and not under equal-gain power-wise combining employed in [1]

The large average SNR asymptote is

$$\overline{P_{e,1}} \approx \frac{1}{2\gamma_0}, \quad \overline{P_{e,2}} \approx \frac{4}{\gamma_0^2} \tag{10}$$

which is the modified form of eq. (41) in [1]; eq. 42 does not require modifications. For m=2 system employing the ZF-MRC weights, eq. (43) and (44) in [1] has to be modified to

$$\overline{P_{e,1}} = 2\overline{P_{BFSK}^{(n-1)}}(\gamma_0) - \overline{P_{BFSK}^{(n-1)}}\left(\frac{\gamma_0}{2}\right) + \Delta P_1(\gamma_0)$$

$$\overline{P_{BFSK}^{(n)}}(\gamma_0) = \frac{1}{2}\left(\frac{2}{2+\gamma_0}\right)^n, \quad \Delta P_1(\gamma_0) = \frac{\gamma_0}{4+\gamma_0}\sum_{i=n-1}^{2n-3}\alpha_i\left(\frac{4}{4+\gamma_0}\right)^i, \quad \alpha_i = \frac{i!a_i}{2} \tag{11}$$

$$\overline{P_{e,2}} = \overline{P_{MRC}^{(n)}}\left(\frac{\gamma_0}{2}\right) - \Delta P_2(\gamma_0), \quad \Delta P_2(\gamma_0) = \frac{1}{2}\frac{\gamma_0}{4+\gamma_0}\sum_{i=n}^{2n-2}\beta_i\left(\frac{4}{4+\gamma_0}\right)^i, \quad \beta_i = \frac{i!b_i}{2^i} \tag{12}$$

where $\overline{P_{BFSK}^{(n)}}(\gamma_0)$ is the average BER of BFSK with n-th order MRC. Their asymptotic behavior is (modified eq. 46 in [1])

$$\overline{P_{e,1}} \approx \frac{1}{2}\left(\frac{1}{\gamma_0}\right)^{n-1}, \quad \overline{P_{e,2}} \approx \left(\frac{2}{\gamma_0}\right)^n \tag{13}$$

Comparing to the (n-1)-th and n-th order MRC BFSK average BER respectively, $\overline{P_{BFSK}^{(n-1)}} \approx (2/\gamma_0)^{n-1}/2$, $\overline{P_{BFSK}^{(n)}} \approx (2/\gamma_0)^n/2$, we conclude that, similar to the outage probability behavior, the effect of the optimal ordering is 3 dB SNR increase in 1st step and doubling of BER at the 2nd step. For large average SNR, the average BLER is dominated by the 1-st step average BER since $\overline{P_{e,1}} \gg \overline{P_{e,2}}$,

$$\overline{P_B} \approx \overline{P_{e,1}} \approx \frac{1}{2}\left(\frac{1}{\gamma_0}\right)^{n-1} \tag{14}$$

For coherent BPSK, $P_e(\gamma) = Q(\sqrt{2\gamma})$ and, for 2x2 system, the average BER is given by (modified eq. (50) in [1]),

$$\overline{P_{e,1}} = \frac{1}{2} - \sqrt{\frac{\gamma_0}{1+\gamma_0}} + \frac{1}{2}\sqrt{\frac{\gamma_0}{2+\gamma_0}}\left(1 + \frac{1}{4(2+\gamma_0)}\right) \tag{15}$$

$$\overline{P_{e,2}} = \frac{1}{2} - \frac{1}{2}\sqrt{\frac{\gamma_0}{2+\gamma_0}}\left(1 + \frac{1}{2+\gamma_0} + \frac{3}{4(4+\gamma_0)^2}\right) \tag{16}$$

They asymptotic behavior is (modified eq. 51 in [1]),

$$\overline{P_{e,1}} \approx \frac{1}{8\gamma_0}, \quad \overline{P_{e,2}} \approx \frac{3}{8\gamma_0^2} \tag{17}$$

In the general case of nx2 system employing the ZF-MRC weights and BPSK modulation, the average BER is given by

$$\overline{P_{e,1}} = 2\overline{P}_{BPSK}^{(n-1)}(\gamma_0) - \overline{P}_{BPSK}^{(n-1)}\left(\frac{\gamma_0}{2}\right) + \Delta P_1(\gamma_0)$$

$$\overline{P}_{BPSK}^{(n)}(\gamma_0) = \frac{1}{2} - \frac{1}{2}\sqrt{\frac{\gamma_0}{1+\gamma_0}}\sum_{i=0}^{n-1}\frac{C_{2i}^i}{4^i(1+\gamma_0)^i}, \tag{18}$$

$$\Delta P_1(\gamma_0) = \frac{1}{2}\sqrt{\frac{\gamma_0}{2+\gamma_0}}\sum_{i=n-1}^{2n-3}\frac{\sigma_i}{(2+\gamma_0)^i}, \quad \sigma_i = (2i-1)!!a_i$$

$$\overline{P_{e,2}} = \overline{P}_{BPSK}^{(n)}\left(\frac{\gamma_0}{2}\right) - \Delta P_2(\gamma_0),$$

$$\Delta P_2(\gamma_0) = \frac{1}{2}\sqrt{\frac{\gamma_0}{2+\gamma_0}}\sum_{i=n}^{2n-2}\frac{d_i}{(2+\gamma_0)^i}, \quad d_i = \frac{(2i-1)!!b_i}{2^i} \tag{19}$$

where $\overline{P}_{BPSK}^{(n)}(\gamma_0)$ is the average BER of BPSK with n-th order MRC, and $C_n^k = n!/(k!(n-k)!)$ is the binomial coefficient. For large average SNR,

$$\overline{P_{e,1}} \approx \frac{C_{2n-3}^{n-1}}{(8\gamma_0)^{n-1}}, \quad \overline{P_{e,2}} \approx \frac{2C_{2n-1}^n}{(4\gamma_0)^n} \tag{20}$$

Similarly to non-coherent BFSK, the effect of optimal ordering is equivalent to a 3 dB SNR increase in 1$^{st}$ step and doubling the BER at the second. For large average SNR, the average BLER is dominated by the 1$^{st}$ step BER,

$$\overline{P_B} \approx \overline{P_{e,1}} \approx \frac{C_{2n-3}^{n-1}}{(8\gamma_0)^{n-1}} \qquad (21)$$

Finally, we note that for other coherent and non-coherent modulation format with BER given by $P_e(\gamma) = \beta Q(\sqrt{\alpha\gamma})$ and $P_e(\gamma) = \beta e^{-\alpha\gamma}$ respectively, the average BER can be evaluated using (11), (12), (18), (19) as $\beta\overline{P_{e,i}}(\alpha\gamma_0/2)$ (coherent mod.) and $2\beta\overline{P_{e,i}}(2\alpha\gamma_0)$ (non-coherent).